\documentstyle[12pt,a4wide]{article}

\renewcommand{\kappa}{{k}}
\newcommand{\ba}{\begin{eqnarray}}
\newcommand{\ea}{\end{eqnarray}}

\begin{document}

\begin{titlepage}

\rightline{hep-th/0503243} \rightline{March 2005}

\begin{centering}
\vspace{2cm}
{\large {\bf Induced Curvature in Brane Worlds by \\ \vspace{2mm}
Surface Terms in String Effective Actions with \\
\vspace{2.5mm} Higher-Curvature Corrections }}

\vspace{1.5cm}

{\bf Nick E.~Mavromatos}$^{a,~*}$  {\bf and} {\bf Eleftherios~Papantonopoulos}$^{b,~**}$  \\
\vspace{3mm}
 $^a$ Physics Department, Theoretical Physics,\\
King's College
London, \\ London WC2R 2LS, U.K. \\
\vspace{3mm}
 $^{b}$ National Technical University of Athens,\\
Physics Department,\\ Zografou Campus, GR 157 80, Athens, Greece.

\end{centering}
\vspace{1.5cm}

\begin{abstract}

In string-inspired effective actions, representing the low-energy
bulk dynamics of brane/string theories, the higher-curvature ghost-free
Gauss-Bonnet combination is obtained by local field redefinitions
which leave the (perturbative) string amplitudes invariant. We
show that such redefinitions lead to surface terms which induce
curvature on the brane world boundary of the bulk spacetime.

\end{abstract}

\vspace{3cm}
\begin{flushleft}
$^{*}~~$ email address: nikolaos.mavromatos@kcl.ac.uk\\
$^{**}~$ email address: lpapa@central.ntua.gr \\

\end{flushleft}
\end{titlepage}

\newpage

\section{Introduction}

In recent years brane world models attracted a lot of attention of
the physics community, as a result of their elegant and novel ways
of tackling fundamental issues of particle physics, such as new
attempts to solve the hierarchy problem~\cite{randall}, and new
approaches to cosmology~\cite{RScosmology}. By representing our
observable world as a thin membrane (brane), in a
higher-dimensional spacetime called bulk, one obtains an
unconventional way of looking at the relation between gravity and
matter. The latter is confined on the brane world, not allowed to
propagate on the bulk space, while gravity is free to propagate on
the bulk, although in warped geometries~\cite{randall},
localization of gravity on the brane world does occur, in the
sense of a peak of the corresponding wavefunction for the graviton
mode~\cite{Garriga:1999yh}.

Such brane world scenaria appear attractive from various view
points. One is the novel mechanisms these models provide in
connection with a possible geometric resolution of the hierarchy
problem~\cite{AADD}, which is represented as a consequence of a
certain distance scale of two brane worlds. The other point of
view is that of cosmology, since the brane world scenaria provide
a variety of new ways of looking at the Physics of the Early
Universe, including inflation, and in general cosmological
evolution~\cite{binetruy}.

  Other important features of these
scenaria include the so-called AdS/CFT correspondence. Indeed, in
most of the scenaria discussed in the literature so far, the bulk
space appears to be anti-de-Sitter (AdS), with negative bulk
cosmological constant. In such models the brane worlds constitute,
then, the (single) boundary of the AdS space, and the holographic
Maldacena conjecture~\cite{maldacena} is in operation: all the
information included in the correlation functions of the matter
quantum field theory on the boundary of such a space can be
obtained by means of a classical (super)gravity theory in the bulk
space, which may be the limit of some underlying
string/brane/M-theory. In this way, one obtains a holographic
image of the bulk space on the boundary, and no information loss
could occur, even if the bulk geometry contains singularities,
such as black holes {\it etc}. The AdS/CFT correspondence
certainly provides a completely novel way of dealing with matter
quantum field theories on brane worlds. A manifestation of this
approach concerns the induced gravity on the
brane~\cite{dvali,Dvali:2000hr} boundary, which results in a
curved spacetime boundary quantum field theory. The AdS/CFT
correspondence on the other hand, on more phenomenological
grounds, may help us, among other things, to understand old
unresolved problems of cosmology like the exit from inflation and
reheating~\cite{Papantonopoulos:2004au}.

A very interesting feature of such brane world models is the fact
that the gravitational scale on the brane world is not a
fundamental scale of the theory, but it is induced in a variety of
ways, and is model dependent. The higher-dimensional bulk
gravitational scale is viewed as the fundamental scale, while the
four-dimensional Planck scale is defined as the coefficient of the
four-dimensional Einstein (scalar curvature) term in the
low-energy effective action, describing the induced gravity on the
brane world~\cite{dvali,Dvali:2000hr}, even if there are curvature
correction terms in the bulk, like Gauss-Bonnet
terms~\cite{rizos}. It is important to notice that the precise
relation of the four-dimensional Planck scale and the bulk
gravitational scale is model dependent, in particular it depends
crucially on the form of the higher-dimensional metric which is a
solution of the bulk gravitational equations of motion.

A brane world model that combines both curvature corrections in
the five-dimensional action, a Gauss-Bonnet term in the bulk and a
scalar curvature term on the brane boundary, was proposed
in~\cite{Kofinas:2003rz}. An important constraint in such kind of
models is the so-called matching conditions of the metric, and
other fields propagating in the bulk, at the discontinuous regions
(``junctions'') of the bulk space defined by the positions of the
brane worlds. By requiring smooth functions of these fields at the
junctions implies important constraints (Israel boundary
conditions), which are quite restrictive for the solutions of the
low-energy theory~\cite{rizos,israel,Charmousis:2002rc}. Recently
it has been pointed out~\cite{Nojiri:2004bx} that the matching
conditions possess ambiguities, in the sense that one may find
consistent conditions for arbitrary boundary gravitational
actions, i.e. $S_1 = \int_{\rm brane} f(R^{(b)})$, where $f(...)$
is an arbitrary function of the boundary scalar curvature
$R^{(b)}$. Such an arbitrariness persists to all orders in
perturbation theory, and includes higher-order curvature
corrections in the effective action. This results in the loss of
the predictability of the model, as far as the low-energy
phenomenology or cosmology are concerned, and thus brings up the
necessity of an underlying fundamental theory, which would
restrict the form of the boundary and bulk actions.

In this work  we will show that strings is a relative simple and
natural framework, where such a restriction occurs.  Starting with
a five-dimensional effective string action including also the
Gauss-Bonnet combination of higher-curvature ghost-free terms, and
no boundary terms in the first place,  we show,  that local field
redefinitions which leave the (perturbative) string amplitudes
invariant, lead to surface terms which induce curvature on the
brane world boundary of the bulk spacetime. These terms introduce
an energy  scale on the brane which depends on the parameters of
the bulk. Such terms of course can be generated also by quantum
corrections of matter fields on the boundary~\cite{dvali}, but we
shall not deal with such corrections here

\section{String Amplitudes in the Bulk, Field-Redefinition Ambiguities and Brane Effective
Actions}

The string theory requires that the effective low-energy action,
in both the bulk and the brane world, admits a systematic
derivative expansion in powers of $\alpha ' p^2$, with $p$ a
generic momentum scale, in the closed string sector, and a
corresponding expansion of $\sqrt{\alpha'}p$ in the open string
sector on the brane world, where $\alpha ' = 1/M_s^2$, with $M_s$
the string scale (which is sometimes called bulk gravitational
scale). It is important to notice that we use the same string
theory, in bulk and brane, in the sense of having a single string
coupling and scale. We also use a systematic derivative expansion
in both the brane and the bulk. This implies that if we restrict
ourselves to quadratic curvature corrections in the low-energy
effective action in the bulk, we shall do the same on the brane
world as well, since we used momentum scales such that
$\sqrt{\alpha'} p \ll 1$.

In the bulk this leads to effective actions which contain a
derivative expansion up to the desired order. In the gravitational
sector, for instance, this leads to several powers of curvature
tensors. In the present work we shall work up to quadratic power
in such curvature tensors, which is equivalent to keeping up to
four derivatives in other terms. For our purposes we shall
restrict ourselves to the graviton and dilaton fields of the
(bosonic part of) the gravitational multiplet of the bulk
(super)string. There is an important result in such perturbative
calculations, pertaining to ambiguities in several coefficients of
such an effective action~\cite{equivth2,equivth}, as a result of local
field redefinitions (i.e. redefinitions involving positive (or
zero) powers of local fields), which leave the scattering
amplitudes invariant.

We consider, for definiteness, the case in which the action is in
five spacetime dimensions. Some remarks are in order at this
point. From a formal point of view, one may think~\cite{verlinde}
of the (bulk) fifth dimension in the spacetime as a (spacelike)
Liouville mode~\cite{ddk}. A more conventional (and probably
safer) approach, which we shall adopt here, is to assume initially
a ten-dimensional spacetime, in which three branes are embedded.
In the bulk one may, then, consider the propagation of closed
strings only, but take the case in which all but one of the bulk
coordinates are compactified. In that case, the induced string
theory amplitudes will formally correspond to those living in an
effective five-dimensional spacetime, in the sense that one may
consider string backgrounds that depend only on the uncompactified
coordinates, and restrict oneself to effective string amplitudes
(or, equivalently, $\sigma$-model conformal-invariance conditions)
for those degrees of freedom.

The low-energy effective action we consider is of the form
 \ba S= S_5 + S_4 \label{s5s4}~, \ea
where $S_5$ denotes bulk contributions, and $S_4$ boundary
contributions on four-dimensional (three space) boundary domain
walls. The four-dimensional part $S_4$ of the action (\ref{s5s4})
is defined as \ba
 S_4 = \sum_{i} \int d^4x \sqrt{-g_{(4)}} e^{\omega \Phi} v(z_i)
\label{s4}~, \ea where $\omega$ is a constant, $\Phi$ is the
dilaton field and \ba g_{(4)}^{\mu\nu}=\left\{
\begin{array}{l}
g^{\mu\nu}\  \,\,\mu,\nu<5\\
0\, \ \ \ \,\ \mbox{otherwise}~,\\
\end{array}\right.
\ea and the sum over $i$ extends over D-brane walls located at
$z=z_i$ along the fifth dimension. The quantity $v(z_i)$ is the
brane tension, which arises also from quantum loop corrections of
matter fields localized on the brane.

The five-dimensional ${\cal O}(\alpha ')$ ($\alpha '$ the Regge
slope) effective action for graviton and dilaton bulk fields
reads: \ba S_5 &=& \int d^5x \sqrt{-g} \left[ -R
-\frac{4}{3}\left(\nabla_\mu \Phi \right)^2 + f(\Phi) \left(\gamma
R^2 + \beta R_{\mu\nu}^2 +
\alpha R_{\mu\nu\rho\sigma}^2\right) \right.\nonumber \\
&~& + \xi(z) e^{\zeta \Phi} + \left. c_2~f(\Phi)\left(\nabla _\mu
\Phi \right)^4 + \dots \right]\ , \label{actionGB} \ea where Greek
indices are five-dimensional indices, and the dots denote other
types of contraction of the four-derivative dilaton terms; these
will not be of interest to us here, for reasons that will be
explained below. The quantity $\xi (z)$ denotes the bulk
cosmological constant, which for simplicity is taken to depend
only on the fifth coordinate $z$. We also have \ba \alpha =+1,
~~f(\Phi)=\lambda ~e^{\theta\Phi}~,~ \lambda =\alpha '/8g_s^2
> 0\ , \label{lambdastring} \ea where $g_s$ is the string
coupling, and $\zeta=-\theta=4/\sqrt{3(D-2)}$~($=4/3$ in five
dimensions of (formal) interest to us here). Moreover, as
mentioned above, in (perturbative) string theory one has the
freedom~\cite{equivth} to redefine the graviton and dilaton fields
so as to ensure that the quadratic-curvature terms in
(\ref{actionGB}) are of the ghost-free GB form~\cite{zwiebach}
\begin{equation}
{\cal R}^2_{GB}=R_{\mu\nu\rho\sigma}R^{\mu\nu\rho\sigma} -
4~R_{\mu\nu}R^{\mu\nu} + R^2\ . \label{GBeq}
\end{equation}
This field-redefinition ambiguity also allows us to consider the
four-derivative dilaton terms in (\ref{actionGB}) as having the
single structure exhibited above. Matching with tree-level string
amplitudes to ${\cal O}(\alpha ')$ then requires~\cite{equivth}\ba
c_2 = \frac{16}{9}\frac{D-4}{D-2}~. \label{tseyt} \ea

The pertinent field redefinitions read (for the normalization of
the field as they appear in (\ref{actionGB})) \ba  g_{\mu\nu} \to
g_{\mu\nu}' &=& g_{\mu\nu} + \alpha ' e^{4\Phi/\sqrt{3(D-2)}}
\Big{(} b_1 R_{\mu\nu} + b_2g_{\mu\nu}R + b_3
\partial_\mu \Phi \partial_\nu \Phi +  \nonumber \\
   &+& b_4 g_{\mu\nu} (\partial \Phi)^2 + b_5 g_{\mu\nu} \Box \Phi
\Big{)}, \nonumber\\   \Phi \to \Phi ' &=& \Phi +
e^{4\Phi/\sqrt{3(D-2)}}\left(c_1 + c_2(\partial \Phi)^2 + c_3 \Box
\Phi\right)~,\label{redef}  \ea where $\Box$ denotes the covariant
D'Alembertian with respect to the old metric $g_{\mu\nu}$, and
$b_i$,$c_i$ are real constant coefficients. For our purposes, for
simplicity and without loss of generality, we shall assume that
the initial bulk action has the form (\ref{actionGB}), that is,
not necessarily ghost free form, but with a single structure of
dilaton four-derivative terms, parameterized by the coefficient
$c_2$. This restricts the class of redefinitions we shall discuss
in the present article to those involving $b_1$,$b_2$ coefficients
only.

The corresponding change of the bulk action (\ref{actionGB}) under
such redefinitions may be summarized as follows: the coefficients
$\beta$ and $\gamma$ change as \ba \beta \to \beta' = \beta -
b_1~, \quad \gamma \to \gamma' = \gamma + \frac{1}{2}\left(b_1 +
(D-2)b_2\right)~, \ea while there are induced also some total
derivative terms of the form (in what follows by $``\delta"$ we
denote the corresponding change in the field under a redefinition
(\ref{redef})) \ba && \delta S_{\rm total~deriv} = M_s^{D-2}\int
d^Dx \sqrt{-g}\lambda \left(\Box g^{\alpha\beta}\delta
g_{\alpha\beta}
- \nabla^\nu \nabla^\mu \delta g_{\mu\nu}\right) = \nonumber \\
&& M_s^{D-2}\lambda ( \frac{b_1}{2} + (D-1)b_2)\int d^Dx
\sqrt{-g}\Box
\left(e^{-\frac{4\Phi}{\sqrt{3(D-2)}}}R\right)~,~~~\lambda
=\frac{\alpha'}{8g_s^2}~. \label{boundary} \ea Such boundary terms
are usually dropped, but in our case, where there are domain wall
boundaries (branes) in the bulk geometry the terms
(\ref{boundary}) may play an important r\^ole in inducing
four-dimensional Einstein curvature terms, as well as cosmological
constant (``brane tension'') contributions on the brane
boundaries. This will be the topic of the next section, where such
terms will be analysed in some detail.

\section{Warped Geometries, Ambiguities and Induced Curvature and
Brane Tension on the Boundaries}

In the five-dimensional bulk case, with brane boundaries, it has
been shown in \cite{rizos} that the Randall-Sundrum (RS) warped
geometries \cite{randall} were solutions of string inspired
effective field theories of the type (\ref{actionGB})
\begin{equation}
ds^2 = e^{-2\sigma(z)}g^{(4)}_{ij}(x)dx^i dx^j + dz^2~,
\label{warpmetric} \end{equation} with \ba \sigma(z)=\sigma_0 + k
|z|~, \label{rssol} \ea where $\sigma_0$ a constant, $x_i$,
$i=0,\dots 3$, are the four-dimensional spacetime coordinates,
$g^{(4)}_{ij}(x)$ is the four-dimensional metric, depending only
on the four-dimensional coordinates, and $z$ is the extra (bulk)
fifth dimension. The presence of the dilaton fields (even if they
were constant in the solution) and of the higher-curvature
corrections in the action (\ref{actionGB}) implied the important
restriction that the Randall-Sundrum parameter $k$ appearing in
the warp factor of the metric, is proportional to the string
coupling $g_s$, with a proportionality factor of order one. The
string coupling was determined by the exponential of the
(constant) dilaton field, $g_s=e^{\Phi_0}$. It was shown in
\cite{rizos} that this was an exact solution of the Gauss-Bonnet
corrected gravitational theory (\ref{actionGB}). Other exact
solutions, include dilatonic domain walls,  in which there are
bulk logarithmic singularities in the metric and the dilaton
field, namely the warp factor and the dilaton acquire the form
\begin{eqnarray}
\sigma (z) = \sigma_0 + \sigma_1 {\rm ln}|1 - \frac{z}{z_s}|~,
\qquad \Phi (z) = \phi_0 -\frac{3}{2} {\rm ln}|1 -
\frac{z}{z_s}|~, \label{domainwalls}
\end{eqnarray}
where $z_s$ the position of the naked singularities, and
$\sigma_1$ a numerical constant. The requirement of finiteness of
the Planck mass and induced cosmological constant on the brane,
required certain restrictions on the domain of
$\sigma_1$~\cite{rizos}.

The presence of domain walls restricts dynamically the available
bulk space. In the simplest scenario, we shall discuss below, the
observable world is a brane which is viewed either as a single
boundary of a bulk anti-de-Sitter space time, which arises
naturally as a solution of the relevant equations of motion, or as
one of the two boundaries of the bulk space, the other being the
domain wall singularity.

In either case, the boundary contributions from the bulk field
redefinitions (\ref{boundary}) result in induced curvature terms
and brane tension contributions due to the following observation:
Observers on brane worlds will have to integrate over the
coordinate $z$ in order to obtain the effective four-dimensional
action. The integrated coefficients of the $R^{(4)}(x)$ terms
yield contributions to the four-dimensional mass scale squared,
$M_4^2$, whilst the rest of the terms contribute to the effective
four-dimensional vacuum energy. Using the warped five-dimensional
metric (\ref{warpmetric}) one obtains \ba\label{r1}
\sqrt{-g}\,R(x)=
\sqrt{-g^{(4)}(x)}\,\left(e^{-2\sigma(z)}\,R^{(4)}(x)
+e^{-4\sigma(z)}\,{\cal R}\right)~,  \ea where \ba {\cal
R}&=&4\,(5\,\sigma'(z)^2-2\,\sigma''(z)) \ea and the superscript
$(4)$ denotes four-dimensional quantities, evaluated on the brane
worlds.

On the other hand, the Gauss-Bonnet curvature combinations yield
terms of the form~\cite{rizos} \ba \lambda e^{-\frac{4}{3}\Phi(z)}
\sqrt{-g}\,R_{GB}(x)&=& \sqrt{-g^{(4)}(x)}\,\lambda
e^{-\frac{4}{3}\Phi (z)}\, \left(4\,e^{-2\sigma(z)}(3\sigma
'(z)^2-2\sigma ''(z))\,
R^{(4)}(x)\right.\nonumber\\
&~&\left.-2\,(R_{\mu\nu\rho\sigma}^2-R_{\mu\nu}^2)+e^{-4\sigma(z)}\,{\cal
R}_{GB}\right) \label{r2}~, \ea where \ba {\cal
R}_{GB}&=&24\,\left( 5\,{{\sigma '(z)}^4} -
    4\,{{\sigma '(z)}^2}\,\sigma ''(z)
     \right)~,
\ea while dilaton derivative terms in the bulk action contribute
\ba \sqrt{-g}\,(\nabla_\mu\Phi)^2=\sqrt{-g^{(4)}(x)}\,\left(
e^{-4\sigma(z)}\,\Phi'(z)^2+e^{-2\,\sigma(z)}\,(\nabla_i\Phi^{(4)}(x))^2\right)
\label{pp1}~. \ea In our work here we shall ignore
four-dimensional dilaton contributions on the brane world for
brevity.

We now observe that, due to (\ref{r1}), the boundary contributions
(\ref{boundary}), arising from the bulk field redefinitions which
cast the bulk action in a Gauss-Bonnet form, yield \ba  \delta
S_{\rm boundary}&=&  M_s^3 \lambda \left(\frac{b_1}{2} +
(D-1)b_2\right)\int d^4x \sqrt{-g^{(4)}(x)}\nonumber \\ &&
\left([-2\sigma'(0)-\frac{4}{3}\Phi'(0)]e^{-2\sigma(0)-\frac{4}{3}\Phi(0)}
R^{(4)}(x)  \right. \nonumber \\
&& + \left. 8e^{-4\sigma(0)- \frac{4}{3}\Phi(0)}\left(-10
(\sigma'(0))^3 + 9\sigma'(0)\sigma''(0)
-\sigma'''(0)\right)  \right. \nonumber \\
&& - \left. \frac{16\Phi '(0)}{3}e^{-4\sigma(0)-
\frac{4}{3}\Phi(0)}(5\,\sigma'(z)^2-2\,\sigma''(z))\right)~,
\label{boundindcurv}\ea from which we see that there are
contributions to the induced four-dimensional curvature and the
brane tension both proportional to the $b_{1}$ and $b_{2}$
coefficients.

In our case, by performing the appropriate field redefinitions
(\ref{redef}) with $b_1, b_2 \ne 0$ only, we arrive at a
 Gauss-Bonnet form of the curvature square terms, which themselves yield
additional contributions (\ref{r2}) to the induced curvature on
the brane $R^{(4)}$, as well as to the brane tension. From the
combination of such terms and (\ref{r1}), one obtains the
expression for the four-dimensional mass $M_4$, as perceived by an
observer living on the brane world located at (finite) $z$
\ba\label{planckexpr} &~& M_4^2 = M_s^3 \int _{-\infty}^{\infty}
dz  e^{-2\sigma (z)} \left(1 - 4~\lambda e^{-\frac{4}{3}\Phi (z)
}(3(\sigma '(z))^2 - 2 \sigma
''(z))\right) + \nonumber \\
&~& M_s^3\lambda \left(\frac{b_1}{2} +
(D-1)b_2\right)[-2\sigma'(0)-\frac{4}{3}\Phi'(0)]
e^{-2\sigma(0)-\frac{4}{3}\Phi(0)}~, ~~~\lambda =
\frac{1}{8M_s^2g_s^2}~. \ea where the last line includes the
pertinent field-redefinition-induced boundary contributions
(\ref{boundindcurv}). It is understood, that the requirement of
positivity of the left hand side of the above equation leads to
non trivial restriction in the range of the various parameters
involved.

In this framework, the four-dimensional effective vacuum energy on
the observable brane world $\Lambda_{\rm total}$ receives two
kinds of contributions: (i) from the tension of the brane world we
are living on, located, say, at $z=z_i=0$, $V_{\rm brane}(z_i)
\equiv e^{\omega \Phi (z_i)}v(z_i)$, and (ii) from the bulk terms
in the action (\ref{s5s4}), that include the cosmological constant
$\xi$, the dilaton derivative terms, as well as the ${\cal R}$,
${\cal R}_{GB}$ dependent terms in (\ref{r1}),~(\ref{r2}).
Therefore, the expression for the total four-dimensional vacuum
energy, including the appropriate boundary contribution
(\ref{boundindcurv}), reads: \ba\label{4dve}
 &~& \Lambda_{\rm total} (0)  = \Omega + V_{\rm brane}(0)~, \nonumber \\
&~& \Omega = \int_{-\infty}^{+\infty} dz\, e^{-4\sigma (z)}
\left[{\vrule height 13pt width 0pt}\xi e^{\frac{4}{3}\Phi (z)}
-\frac{4}{3} (\Phi '(z))^2 - 20(\sigma '(z))^2  + 8 \sigma
''(z)\right.
+ \nonumber \\
&~&\left. \lambda e^{-\frac{4}{3}\Phi(z)}\left(24~(5(\sigma
'(z))^4 - 4~(\sigma '(z))^2\sigma ''(z)) + c_2 (\Phi '(z))^4
\right){\vrule height 13pt width 0pt}\right],
\nonumber \\
&~& V_{\rm brane}(0)=e^{\omega \Phi(0)}v_i + \nonumber \\
&~& 8M_s^3\lambda \left(\frac{b_1}{2} + (D-1)b_2\right)
e^{-4\sigma(0)- \frac{4}{3}\Phi(0)}\{\left(-10 (\sigma'(0))^3 +
9\sigma'(0)\sigma''(0)
-\sigma'''(0)\right) - \nonumber \\
&& \frac{16\Phi '(0)}{3}(5\,\sigma'(z)^2-2\,\sigma''(z))\}~. \ea

In physically acceptable situations, the quantities $M_4^{2}$ and
$\Lambda_{\rm total}$ should be finite, which, in the case in
which one encounters bulk singularities, implies certain
integrability conditions, as discussed in \cite{rizos}. This is an
important restriction on model building. For the logarithmic
solutions (\ref{domainwalls}), for instance, the above
requirements, when applied to the $b_i$-independent parts of
$\Lambda$ and $M_4^{2}$ only, imply \ba \sigma_1 < -\frac{1}{4}~.
\label{sigma1} \ea

The dependence of physical quantities, such as brane vacuum
energies and the four-dimensional mass, on redefinition
coefficients gives them a degree of arbitrariness. Indeed, the
combination of field-redefinition coefficients appearing in
(\ref{planckexpr}), would indicate that the relation between the
Planck mass and the string mass (scale) is not fixed once an exact
solution is given, but depends on the redefinition coefficients
$b_{1}$ and $b_{2}$. Moreover, as can be seen from the formulae
above, it is not possible to absorb these coefficients in suitable
redefinitions of constants in the effective action. One therefore
should insist that $b$-dependent terms should define another
energy scale when an exact solution of warped geometry is imposed
not determined by the parameters of the theory. Such a requirement
would in general impose further restrictions on the form of exact
solutions of \cite{rizos}, which is a very interesting feature to
be discussed next.

\section{Induced Curvature on the Brane and the Need for a New Energy Scale }

A brane world model of high curvature terms in the bulk and
induced curvature terms on the brane should be characterized by
the following fundamental parameters: three energy scales, i.e.
the fundamental Planck mass $M_5$ (or bulk scale $M_s$ in our
five-dimensional toy-string model), the induced-gravity crossover
energy scale $r_c^{-1}$, and the Gauss-Bonnet coupling energy
scale $\lambda^{-1/2} = 2\sqrt{2}g_s/\sqrt{\alpha '}$, and two
vacuum energies, i.e. the bulk cosmological constant $\xi$ and the
brane tension $V_{\rm brane}(0)$. These parameters would determine
any physical process on the brane, as its cosmological evolution
\cite{Kofinas:2003rz}.

Had we ignored the surface terms (\ref{boundindcurv}) which are
induced on the brane because of the freedom we have to do field
redefinitions to the graviton and dilaton fields in the bulk, the
Gauss-Bonnet term projected on the brane would give a general form
for the four-dimensional gravity mass scale on the brane
\ba\label{planckexpr4} &~& M_4^2 = M_s^3 \int _{-\infty}^{\infty}
dz  e^{-2\sigma (z)} \left(1 - 4~\lambda e^{-\frac{4}{3}\Phi (z)
}(3(\sigma '(z))^2 - 2 \sigma ''(z))\right)~. \ea If a solution of
the background metric (\ref{warpmetric}) is given and the boundary
conditions on the brane are satisfied then, the above relation
defines the physical four-dimensional Planck mass on the brane in
terms of the five-dimensional mass $M_s$. If the surface terms
(\ref{boundindcurv}) are included there is another contribution to
the four-dimensional mass \ba  M_s^3\lambda \left(\frac{b_1}{2} +
(D-1)b_2\right)[-2\sigma'(0)-\frac{4}{3}\Phi'(0)]
e^{-2\sigma(0)-\frac{4}{3}\Phi(0)}~.\ea Even if the background
solution is given this relation does not fix the relation between
$M_4$ and $M_5$, introducing another scale in the model. If we
define
\begin{equation}
r_{c}=\frac{M_4^2}{M_s^3}=\lambda \left(\frac{b_1}{2} +
(D-1)b_2\right)[2\sigma'(0)+\frac{4}{3}\Phi'(0)]
e^{-2\sigma(0)-\frac{4}{3}\Phi(0)}~\,, \label{crosoever}
 \label{distancescale}
 \end{equation} we recover the energy scale of induced
 gravity, known as crossover scale. The contribution of the
 surface terms to the vacuum energy in relation (\ref{4dve}) can also be
 expressed in terms of
 the crossover scale.

Depending on the specific solution one considers, the expressions
for the crossover scale and brane tension are different. For
instance, in the Randall-Sundrum exact solution~\cite{randall},
the dilaton is a constant, while $\sigma '(0) = k =
1/2\sqrt{3\lambda} \propto g_s / \sqrt{\alpha '}$ as a result of
the higher-curvature corrections~\cite{rizos}. All the higher
$z$-derivatives of $\sigma (z)$ vanish in the Randall-Sundrum
solution. In that case, the crossover scale (and also the brane
tension) depend only  on the linear combination
$\left(\frac{b_1}{2} + (D-1)b_2\right)$ of the ambiguous
coefficients.

On the other hand, in the logarithmic domain-wall solution
(\ref{domainwalls}), we have: $\sigma ´(0)= \sigma_1/z_s$ (and
similarly for its higher $z$-derivatives), and thus one has a
dependence on the position of the domain wall brane $z_s$ as well
as the parameter $\sigma_1$ of the solution.

The dependence of physical quantities, such as the crossover scale
and the induced brane tension on the field-redefinition ambiguous
parameters $b_i$ implies of course a ``breakdown'' of the spirit
(but not the letter) of the equivalence theorem in the case of
branes in the following sense: although the low-energy effective
closed string theory, which lives in the bulk, might enjoy on its
own such redefinition ambiguities, in the sense that it is defined
perturbatively through the respective scattering amplitudes, which
are insensitive to these redefinitions (thus the ``letter'' of the
equivalence theorem holds), nevertheless the presence of brane
boundaries, with non trivial string matter on them, leads -
through boundary contributions- to a dependence of physical
quantities of the brane worlds on these coefficients (thereby
leading to an obvious violation of the ``spirit'' of the theorem).

\section{Conclusions and Discussion}

In this work we have argued that in string theory
curvature terms in the field-theoretic
effective (low-energy) bulk actions
induce scalar curvature and brane tension terms on the boundary.

This happens as a result of the field redefinition ambiguities
that characterise the effective actions constructed out of bulk
closed string-states scattering amplitudes. The ghost-free
Gauss-Bonnet combination, which from a field theoretic point of
view is unique, is not special in string theory, precisely as a
result of the fact that the underlying string model is always
ghost free. In terms of perturbative scattering amplitudes this is
achieved by the fact that one can always perform field
redefinitions that can cast the target-space higher curvature
terms in the effective action in the Gauss-Bonnet combination.

However, in the presence of brane boundaries, this procedure
leaves non-trivial curvature and induced brane tension terms,
which depend on the ambiguous coefficients. Such ambiguities
characterise the crossover scale, as well as the brane tension,
and as such cannot be ignored. On the contrary, they are partially
fixed by relating the resulting crossover scale with the four
dimensional Planck mass, and the induced brane tension with
current values of a positive cosmological constant set by
astrophysical observations.

An interesting issue concerns the explanation of such ambiguities
in the crossover scale from a $\sigma$-model viewpoint. Indeed, it
is well known~\cite{equivth2} that the local field redefinitions
in the bulk string theory correspond to renormalisation-group
scheme changes in the corresponding bulk-closed-string
$\sigma$-model, amounting to adding local counterterms in the
corresponding $\sigma$-model action. Under such scheme changes,
the world-sheet renormalisation-group $\beta$-functions (or rather
the so-called Weyl anomaly coefficients~\cite{equivth2}) change
with a ``Lie-derivative'' in theory space, the latter being
defined by the background $\sigma$-model couplings/fields $g^i$
\ba g^i \to g^i + \delta g^i~, \qquad \beta^i \to \beta^{i´} =
\beta^i + \delta g^j \partial_j \beta^i - \beta^j \partial_j
\delta g^i~, \ea where we used infinitesimal field redefinitions
for brevity (this situation is representative of our situation, in
which the redefinition parameter is provided by the string Regge
slope $\alpha' \to 0$). In ordinary string theory, conformal
invariance, in the sense of the vanishing of the $\beta^i = 0 $,
in a given scheme, is equivalent to string equations of motion
$\delta S[g']/\delta g^{i'}= 0$ in another scheme (denoted by a
prime, $g^{i'}$), in the sense that there is always a ``relative''
scheme choice $\delta g^i$ which guarantee this.

Let us see how this situation is modified in our case. From the
point of view of an open string living on the brane, there is an
open $\sigma$-model describing string excitations of the brane
world. When a bulk closed string crosses the brane boundary, it
may split~\cite{cardy} into two open strings, with their ends
attached on the boundary. From a conformal field theory view point
such a splitting may be described by the approach of a world-sheet
bulk vertex operator $O$ of a closed string, corresponding to,
say, a graviton excitation, to a world-sheet boundary $\partial
\Sigma$ (parametrised by a real coordinate $s$). In such a case
one has a novel operator product expansion~\cite{cardy} \ba
O(z,{\bar z}; s) \sim \sum_{I} (2s)^{\Delta_I - \Delta
_0}C^A_{O,{\cal E}_I}{\cal E}_I(s)~, \ea provided that the set of
boundary conditions $A$ does not break conformal symmetry. Above,
$z, \bar z$ denote bulk world sheet coordinates, $\Delta $ are the
corresponding scaling dimensions of the vertex operators, and
${\cal E}_I$ denotes a complete set of boundary operators. The
splitting amplitudes $C^A_{O,{\cal E}_I}$ can be
expressed~\cite{cardy} in terms of bulk operator product expansion
(o.p.e.) coefficients $c^{i}_{jk}$, appearing in the bulk
$\sigma$-model Weyl anomaly coefficients $\beta^i = y^i g^i +
c^i_{jk}g^jg^k + \dots $ The o.p.e. coefficients appear in the
string effective action $[g]$, since the latter is related to the
off-shell $\beta$-functions by means of a gradient flow relation
\ba \beta^i = {\cal G}^{ij}\frac{\delta S[g]}{\delta g^j}~,\ea
with ${\cal G}^{ij}$ Zamolodchikov's metric in theory space,
related to the two-point function of the corresponding vertex
operators.

From this point of view, a bulk redefinition of $g^i$
affects the
coefficients $c^i_{jk}$, according to what was mentioned above,
and hence it will
affect the splitting amplitudes $C^A_{O,{\cal E}_I}$, thereby inducing
local counterterms in the $\sigma$-model describing the world-sheet
boundary conformal field theory.
In this context, the effective four-dimensional action on the brane world
corresponds to the effective
target-space action of a $\sigma$-model
on the brane propagating on four-dimensional metric backgrounds. In
in view of the above-described
o.p.e. bulk/boundary relations, this effective action
depends in general on the splitting amplitudes, since new
world-sheet short-distance divergencies arise.
This is the root of the violation of the spirit of the equivalence theorem
by means of the brane boundary terms.

Although the bare string tension $1/4\pi\alpha'$ is the same
between the bulk and boundary string theory, nevertheless the
effective four-dimensional target space gravitational scale
appearing in front of the Einstein term in the four-dimensional
target space effective action need not be the same. Indeed, in
view of the above-mentioned induction of local counterterms in the
boundary $\sigma$-model theory, as a result  of renormalisation
group scheme changes in the bulk $\sigma$-model, one would obtain
a redefinition of the four-dimensional gravitational scale,
appearing in front of the Einstein term of the four-dimensional
effective action, as well as in a renormalisation of the brane
tension itself, as we have seen in previous sections.

The appearance of ambiguities in the brane vacuum energy (induced
cosmological constant) brings up another important issue. Namely,
an (open) string in a spacetime with a cosmological constant is
not in general conformal invariant, and thus an appropriate
procedure should be implemented to restore the conformal symmetry.
One way to resolve it is to insist on a  fine tuned cancellation
of the brane cosmological constant by means, e.g. of
supersymmetry. The other, and more natural from our point of view,
is to dress the boundary $\sigma$-model theory by means of a
Liouville field~\cite{ddk}. The spacetime signature of the latter
depends on the signature of the induced vacuum energy.

In some of the examples we discussed, it is possible to chose
families of solutions of the higher curvature gravity such that
the induced vacuum energy on the brane is free from the field
redefinition ambiguities. For instance, for the domain wall
solutions, this can be achieved by requiring the coefficient of
the ambiguous term in the vacuum energy to vanish. This leads to a
cubic algebraic equation appearing in (\ref{4dve}), which has
always a real solution. In fact in this case it can be checked
that this solution also satisfies the physical constrain
(\ref{sigma1}). This would ensure a definite sign for the induced
vacuum energy of the brane, fixed by the Gauss Bonnet
combination~\cite{rizos}, which can then be dealt with either via
Liouville dressing, or through the traditional Fischler-Susskind
mechanism~\cite{sussk}, according to which higher-order topologies
on the world sheet (string loops) of the open strings on the brane
can be held responsible for the appearance of a cosmological
constant. Notice that for the Randall-Sundrum solution, it is not
possible to choose the coefficient of the ambiguous term in the
vacuum energy to vanish.

These considerations cast some doubt on the predictability of
brane effective field theory models, embedded in a string theory
framework. In string theory, the string bulk scale $M_s$ is an
arbitrary parameter, and according to our considerations above,
this seems to be also the case of the four-dimensional induced
brane scale as well as the vacuum energy on the brane world. It
may well be that such fundamental issues are not resolved until a
fully non-perturbative string/brane theory is available.

\section*{Acknowledgments}

We acknowledge informative discussions with Naresh Dadhich and
Alex Kehagias. This work is partially supported by funds made
available by the Empeirikeion Foundation (Greece) and by the Greek
Ministry of Education (Research program ``Pythagoras''). N.M.
wishes to thank the National Technical University of Athens for
hospitality, during which this work was started and  E.P. wishes
to thank the Physics Department of King's College London, for
hospitality during the last stages of this work.

\end{document}